\documentclass[fleqn,twoside]{article}
\usepackage{espcrc2}

\usepackage{dcolumn}
\usepackage{bm}
\usepackage{amssymb}  
\usepackage{amsmath}
\usepackage{epsfig}    
\usepackage{here}

\usepackage{graphicx}
\usepackage[figuresright]{rotating}

\def\beq{\begin{equation}}
\def\eeq{\end{equation}}
\def\bea{\begin{array}}
\def\eea{\end{array}}
\def\be{\begin{equation}}
\def\ee{\end{equation}}
\def\ba{\begin{eqnarray}}
\def\ea{\end{eqnarray}}

\def\to{\rightarrow}

\def\[{\left[}
\def\]{\right]}
\def\({\left(}
\def\){\right)}

\def\mutau{{$\mu N\rightarrow\tau X$~}}

\def\sm0{{\widetilde{m}_0}}

\def\U1em{{U(1)_{\rm em}}}
\def\to{\rightarrow}

\def\sq2{\sqrt{2}}

\def\ee{e^+e^-}

\def\End{\end{document}}

\def\Journal#1#2#3#4{{#1} {\bf #2}, #3 (#4)}

\def\PLB{{\rm Phys. Lett.}  B}

\def\PRL{\rm Phys. Rev. Lett.}
\def\PRD{{\rm Phys. Rev.} D}

\title{Search for lepton flavor violation 
via the intense high-energy muon beam}
\author{S. Kanemura\address[osaka-u]{Department of Physics, 
         Osaka University, Toyonaka, Osaka 560-0043, Japan}, 
         Y. Kuno\addressmark[osaka-u],
     M. Kuze\address{Department of Physics, 
     Tokyo Institute of Technology, 
     Tokyo 152-8851, Japan}, T. Ota\addressmark[osaka-u]}

\begin{document}

\begin{abstract}
A deep inerastic scattering process \mutau is discussed 
to study lepton flavor violation between muons and tau leptons.
In supersymmetric models, 
the Higgs boson mediated diagrams could be important for this reaction. 
We find that at a muon energy ($E_{\mu}$) higher than 50 GeV, 
the predicted cross section significantly increases due to 
the contribution from sea $b$-quarks.
The number of produced tau leptons can be $\mathcal{O}(10^4)$ 
at $E_{\mu}$= 300 GeV from $10^{20}$ muons, whereas 
$\mathcal{O}(10^2)$ events are given at $E_{\mu}= 50$ GeV. 
\vspace{1pc}
\end{abstract}

\maketitle

\section{Introduction} 

In a model based on Supersymmetry (SUSY), 
slepton mixing is a source of Lepton Flavor Violation (LFV). 
Two types of effective LFV couplings are induced at low energies; 
i.e., those mediated by the neutral gauge bosons
and those by the neutral Higgs bosons. 
In contrast to the gauge boson mediation, 
the contributions from the Higgs-mediation do not 
decouple even if the soft SUSY breaking scale 
is as large as $\mathcal{O}(1)$ TeV~\cite{Babu}.
To study the Higgs-mediated LFV couplings  
the tau-associated processes are useful 
because they are proportional 
to the mass of the relevant charged leptons. 
The LFV couplings associated with a tau lepton 
have been measured at $B$ factories by searches 
for rare tau decays, such as  
$\tau \to \mu \gamma$, 
$\tau \to 3 \mu$, 
$\tau \to \mu \pi\pi$, 
$\tau \to \mu \eta$, etc.
At future collider experiments, they are directly tested 
via the decays of the Higgs bosons ($\Phi^{0}$), 
$\Phi^{0} \rightarrow \tau^{\pm} \mu^{\mp}$~\cite{Assamagan,Higgs-TauMu-LC}. 

In this talk, we discuss a \mutau reaction 
at the deep inelastic scattering (DIS) region  
with high-intensity and high-energy muon beams
as an alternative approach of searching 
for the LFV couplings associated with a tau lepton. 
Here, $N$ is a target nucleon, and $X$ represents 
all final state particles. 
Sher and Turan have discussed this process 
in a model independent approach~\cite{Sher-DIS}.  
Instead, we here consider this process in the framework of SUSY~\cite{kkko}.

\section{ The \mutau cross section}

When the scalar LFV coupling of $\tau\mu qq$ 
is independent of the other types of
couplings, its experimental constraint comes from the   
$\tau \to \mu\pi\pi$ result. 
The total cross section of the process \mutau mediated by the 
scalar LFV coupling could then be as large as 0.5 fb at muon energy 
($E_{\mu}$) of 50 GeV. 
For this case, with $10^{20}$ muons per year on a 
$\mathcal{O}(10^2)$ g/cm$^2$ target mass, about 
$10^{6}$ of the \mutau events can be produced, or no observation of the 
\mutau signal would improve the limits by six orders of 
magnitude~\cite{Sher-DIS}.

\begin{figure}
\includegraphics[width=6.4cm]{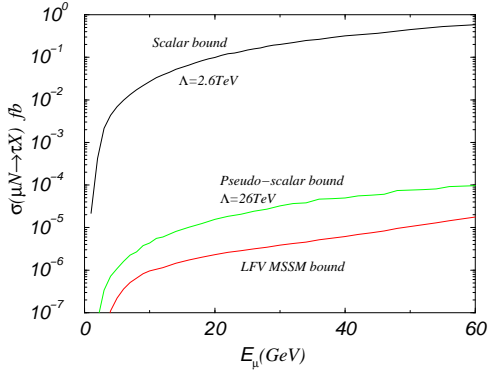}
\caption{%
The upper bounds on the cross section of the $\mu^- N \to \tau^- X$ 
DIS process, assuming the effective scalar and pseudo-scalar 
couplings~\cite{Sher-DIS} and the MSSM Higgs coupling 
constrained from the current data~\cite{kkko}. CTEQ6L is used for the PDF. 
}\label{Fig:tot}
\end{figure}

In the Minimal Supersymmetric Standard Model (MSSM), 
the Higgs boson couplings are related each other. 
In the decoupling region ($m_{A} \gtrsim 150$ GeV),
where $m_{H} \simeq m_{A}$ and $\sin(\alpha-\beta) \simeq -1$,
the scalar coupling $\mathcal{C}^{hH}_L$ is nearly equal to the 
pseudo-scalar coupling $\mathcal{C}^{A}_L$. 
Therefore, both couplings are determined by the more constrained one, 
namely the pseudo-scalar coupling~\cite{Sher-Tau-MuEta}. 
It is constrained by the $\tau \rightarrow \mu \eta$ decay  
(${\rm Br}(\tau \to \mu \eta) 
< 3.4 \times 10^{-7}$)~\cite{Belle-Tau-MuEta}. 
In Fig.~\ref{Fig:tot}, the total cross sections are shown 
for the cases of the effective scalar and pseudo-scalar couplings 
as well as the SUSY model with the maximal value for the couplings 
under the current data of the rare tau decays. 
The largest values of ${\mathcal{C}^{hH}_L}$ 
and ${\mathcal{C}^{A}_L}$ can be realized with
$m_{\text{SUSY}} \sim \mathcal{O}(1)$ TeV and 
the Higgsino mass $\mu \sim \mathcal{O}(10)$ TeV.
It should be noted that 
in such a situation, the gauge boson mediated
couplings are strongly suppressed. 

\begin{figure}
\includegraphics[width=6cm]{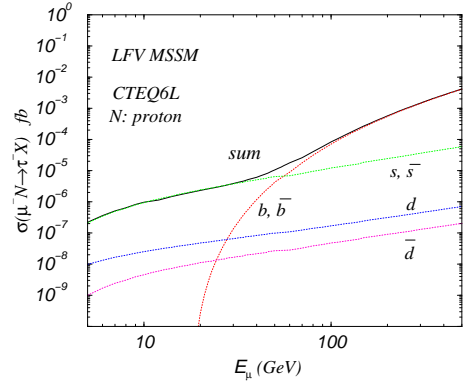}
\caption{%
Cross section of the $\mu^- N \to \tau^- X$ DIS process 
as a function of the muon energy for 
the Higgs mediated interaction. It is assumed that the initial 
muons are purely left-handed.
CTEQ6L is used for the PDF. 
}\label{Fig:total-cross-section-vs-Emu}
\end{figure}

We evaluate the cross sections of the \mutau reaction 
in the DIS region for the Higgs-mediated interaction 
with the maximally allowed values of the effective couplings
for each quark contribution: see Fig.\ref{Fig:total-cross-section-vs-Emu}. 
The cross section sharply increases above $E_{\mu} \sim 50$ GeV.
This enhancement comes from a consequence of the $b$-quark
contribution in addition to the $d$ and $s$-quark contributions. 
The cross section is enhanced by one order of magnitude when the muon
energy changes from 50 GeV to 100 GeV. Typically, 
for $E_{\mu} = 100$ GeV and $E_{\mu} = 300$ GeV, the cross section 
is $10^{-4}$ fb and $10^{-3}$ fb, respectively. 

Next, we study cases where the gauge-boson mediation is dominant; i.e., 
$m_{\text{SUSY}} \sim \mathcal{O}(100)$ GeV. 
Since 
Br$(\tau \rightarrow \mu\gamma)< 3.1\times 10^{-7}$~\cite{Belle-Tau-MuGamma}, 
the contribution from the tensor interaction is found to be 
smaller than that from the Higgs boson mediation 
by about five orders of magnitude. 
On the other hand, the vector and axial-vector interactions 
are suppressed at the same level as the pseudo-scalar 
interaction~\cite{Black-Han-He-Sher}.
Therefore, their contributions can be as large as 
those for the Higgs boson mediation, 
if $E_\mu$ is less than 50 GeV~\cite{Sher-DIS}.   
At higher energies, the cross section for 
the gauge boson mediation becomes much smaller than 
that for the Higgs boson mediation because of 
no enhancement by the $b$-quark sub-process.
\begin{figure*}
{\center
\includegraphics[width=6cm]{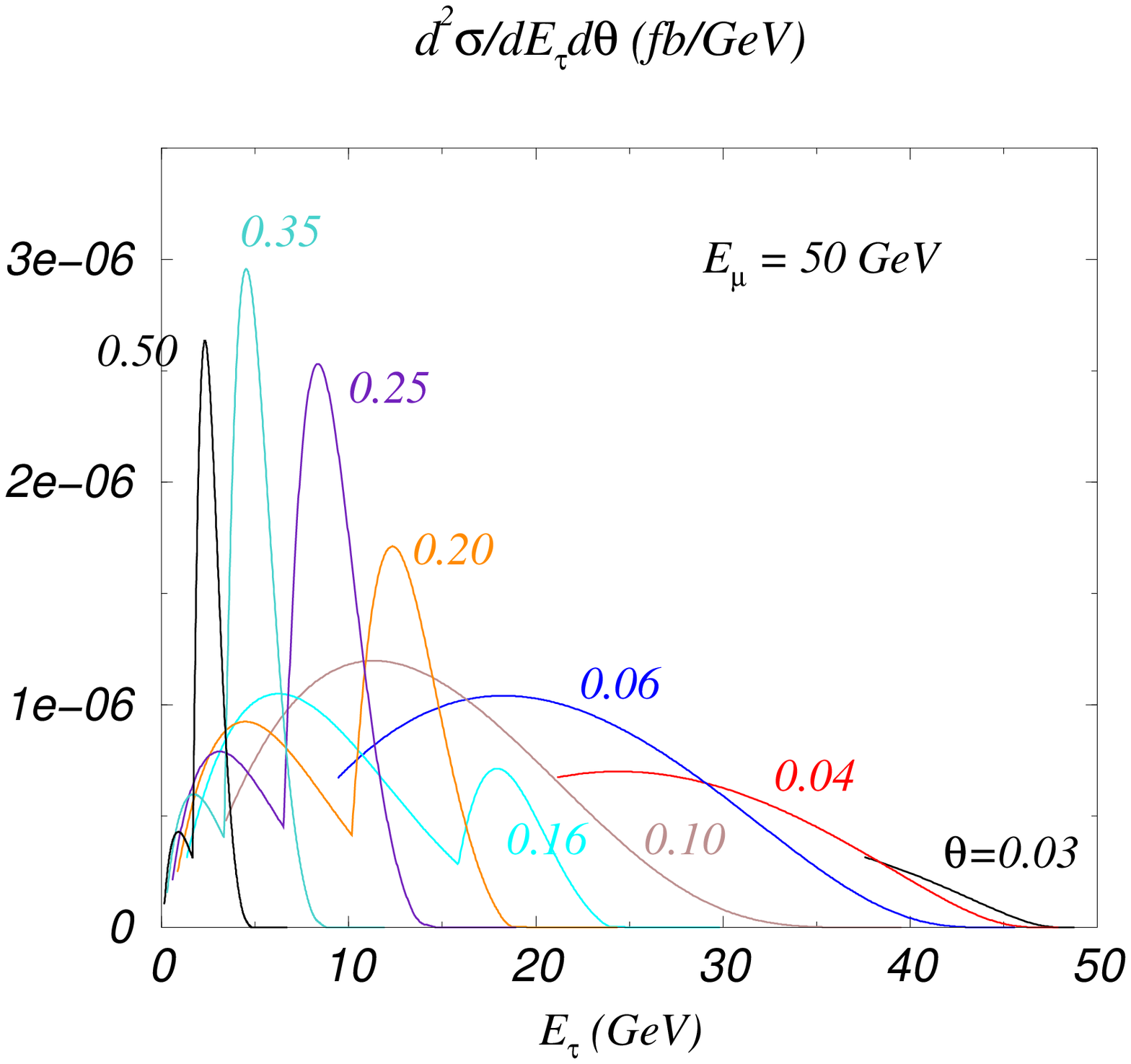}
\hspace{15mm}
\includegraphics[width=6cm]{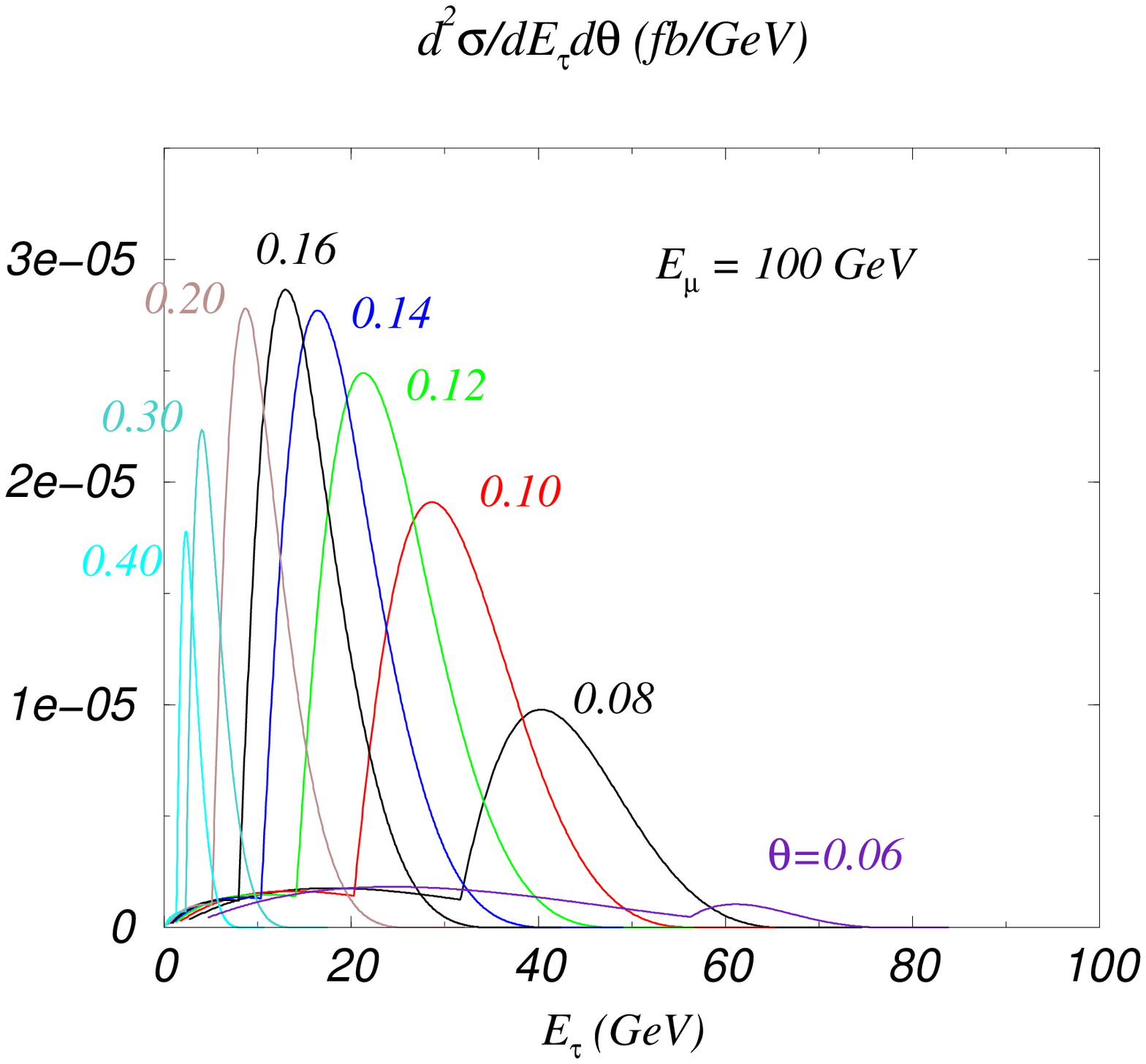}
\caption{%
The differential cross section of the tau 
from the $\mu^- N \to \tau^- X$ DIS process as a function 
of the tau energy ($E_{\tau}$) and the tau emission angle 
($\theta$) with respect to the forward direction 
for $E_{\mu} = 50$ GeV (left) and $E_{\mu} = 100$ GeV (right).
It is assumed that the initial muons are purely left-handed.
}}
\label{Fig:angle-dependence}
\end{figure*}

\section{The \mutau Phenomenology} 

With the intensity of $10^{20}$ muons per year and 
the target mass of 100 g/cm$^2$, 
about $10^4$ ($10^2$) events could be 
expected for $\sigma(\mu N\rightarrow \tau X)=10^{-3} ~(10^{-5})$ fb, 
which corresponds to $E_{\mu}=300$ $(50)$ GeV 
from Fig.~\ref{Fig:total-cross-section-vs-Emu}. 
This would provide good potential to improve the sensitivity by four (two)
orders of magnitude from the present limit from 
$\tau\rightarrow\mu\eta$ decay, 
respectively. Such a muon intensity could be available at a future
muon collider and a neutrino factory. 

In the Higgs boson mediated interaction, the tau leptons 
are emitted at a relatively large angle 
with respect to the beam direction. This is  
in contrast to the gauge mediated interaction where
the tau leptons are forward-peaked.
In Fig.~\ref{Fig:angle-dependence}, the $E_\tau$ dependence 
in the differential cross section is shown for each 
$\theta$ at $E_{\mu}$ = 50 and 100 GeV. 

To identify the tau lepton from the \mutau reaction, direct 
measurement of tau lepton tracks (such as by emulsions) 
might not be possible at such a high beam rate. 
Instead, the identification might be possible 
by tagging the tau decay products and observing their decay kinematics. 
Among various decay modes, one might consider leptonic decays of the 
tau leptons. Another candidate could be 
to detect a hadron from the two-body tau decays.  
The branching ratios, such as 
$\tau \rightarrow \nu_{\tau} \pi$, $\nu_{\tau} \rho$
and $\nu_{\tau} a_{1}$, are about 0.3 in total. 
In particular, in SUSY models with left-handed slepton mixing, 
the $\tau^{-}$($\tau^+$) produced through the Higgs-mediated interaction 
is only right-handed (left-handed) for an incident left-handed 
$\mu^-$ beam (right-handed $\mu^+$ beam).
The hadrons from the right-handed $\tau^{-}$ decay (left-handed $\tau^+$     
decay) tend to be emitted in the direction of the parent tau lepton, 
and therefore be rather energetic. 
Therefore, the signature of the events could be a hard hadron 
at a relatively large angle from the beam direction 
(namely a hadron with large transverse momentum $p_{T}$) 
and some missing energy. 
Those hadrons from the tau decay should be discriminated  
from the hadrons from the target nucleons  
which have mostly soft energies.

\section{Conclusions}

This process can be useful to search for the Higgs-boson mediated 
LFV coupling, especially when $E_\mu$ is higher than 50 GeV. 
There, the contributions from the sea $b$-quarks become significant,  
and the cross section is drastically enhanced.


\begin{thebibliography}{9}

\bibitem{Babu} K.S.~Babu and C.~Kolda, \Journal{\PRL}{89}{241802}{2002};
               A.~Dedes, J.~Ellis, and M.~Raidal, 
               \Journal{\PLB}{549}{159}{2002}; 
               A.~Brignole and A.~Rossi,
               \Journal{\PLB}{566}{217}{2003}.

\bibitem{Assamagan}
K.~Assamagan, A.~Deandrea, and P-A.~Delsart,
\Journal{\PRD}{67}{035001}{2003}.

\bibitem{Higgs-TauMu-LC}
S.~Kanemura, K.~Matsuda, T.~Ota, T.~Shindou, E.~Takasugi, and 
K.~Tsumura, 
\Journal{\PLB}{599}{83}{2004}, hep-ph/0406364. 

\bibitem{Sher-DIS}
M.~Sher and I.~Turan,
\Journal{\PRD}{69}{017302}{2004}.

\bibitem{kkko} S.~Kanemura, Y.~Kuno, M.~Kuze, and T. Ota, 
               to appear in Phys. Lett. B (hep-ph/0410044).

\bibitem{Sher-Tau-MuEta}
M.~Sher,
\Journal{\PRD}{66}{057301}{2002}.


\bibitem{Belle-Tau-MuEta}
Belle Collaboration, Y.~Enari {\it et al.},
\Journal{\PRL}{93}{081803}{2004}.

\bibitem{Belle-Tau-MuGamma}
Belle Collaboration, K.~Abe {\it et al.},
\Journal{\PRL}{92}{171802}{2004};
%

\bibitem{Black-Han-He-Sher}
D.~Black, T.~Han, H-J.~He, and M.~Sher,
\Journal{\PRD}{66}{053002}{2002}. 


\end{thebibliography}
\end{document}